\RequirePackage{lineno}
\documentclass[prd,preprint,tightenlines,superscriptaddress,showpacs]{revtex4}
\usepackage{amssymb}
\usepackage{}
\usepackage{graphicx}
\usepackage{epsfig}
\usepackage{dcolumn}
\usepackage{bm,color}
\usepackage{multirow}
\usepackage{verbatim}

\newcommand{\Br}{{\cal B}}

\newcommand{\piz}{\pi^0}

\newcommand{\etap}{\eta^{\prime}}

\newcommand{\etacp}{\eta_{c}(2S)}
\newcommand{\psp}{\psi(3686)}

\newcommand{\jpsi}{J/\psi}
\newcommand{\ks}{K_{S}^{0}}
\newcommand{\EE}{e^+e^-}

\newcommand{\pp}{\pi^+\pi^-}
\newcommand{\kk}{K^+K^-}

\newcommand{\ra}{\rightarrow}

\newcommand{\kskp}{K^{0}_S K^{\pm}\pi^{\mp}}
\newcommand{\kkp}{K^{+} K^{-}\pi^{0}}
\newcommand{\kskppp}{K^{0}_{S} K^{\pm}\pi^{\mp}\pi^{+}\pi^{-}}

\newcommand{\beq}{\begin{equation}}
\newcommand{\eeq}{\end{equation}}
\newcommand{\bitm}{\begin{itemize}}
\newcommand{\eitm}{\end{itemize}}
\newcommand{\mev}{\mathrm{MeV}}

\newcommand{\mevcc}{\mathrm{MeV}/c^2}
\newcommand{\gev}{\mathrm{GeV}}
\newcommand{\gevc}{\mathrm{GeV}/c}
\newcommand{\gevcc}{\mathrm{GeV}/c^2}

\begin{document}
\linenumbers
\title{\quad\\[1.0cm]\boldmath Evidence for $\etacp$ in $\psp \ra \gamma \kskppp$}

\author{\small
M.~Ablikim$^{1}$, M.~N.~Achasov$^{6}$, O.~Albayrak$^{3}$, D.~J.~Ambrose$^{39}$, F.~F.~An$^{1}$, Q.~An$^{40}$, J.~Z.~Bai$^{1}$, R.~Baldini Ferroli$^{17A}$, Y.~Ban$^{26}$, J.~Becker$^{2}$
% ...
\\
\vspace{0.2cm}
(BESIII Collaboration)\\
\vspace{0.2cm} {\it
$^{1}$ Institute of High Energy Physics, Beijing 100049, People's Republic of China\\
$^{2}$ Bochum Ruhr-University, D-44780 Bochum, Germany\\
$^{3}$ Carnegie Mellon University, Pittsburgh, Pennsylvania 15213, USA\\
% ...
$^{6}$ G.I. Budker Institute of Nuclear Physics SB RAS (BINP), Novosibirsk 630090, Russia\\
% ...
$^{17}$ (A)INFN Laboratori Nazionali di Frascati, I-00044, Frascati, Italy; (B)INFN and University of Perugia, I-06100, Perugia, Italy\\
% ...
$^{26}$ Peking University, Beijing 100871, People's Republic of China\\
% ...
$^{39}$ University of Rochester, Rochester, New York 14627, USA\\
$^{40}$ University of Science and Technology of China, Hefei 230026, People's Republic of China\\
% ...
}
\vspace{0.4cm}}

\begin{abstract}
We search for the M1 radiative transition $\psp \ra \gamma \etacp$ by reconstructing the exclusive $\etacp \ra \kskppp$ decay using 1.06 $\times$ $10^{8}$ $\psp$ events collected with the BESIII detector.
The signal is observed with a statistical significance of greater than $4$ standard deviations. The measured mass of the $\etacp$ is 3646.9 $\pm$ 1.6(stat) $\pm$ 3.6(syst) $\mevcc$, and the width is 9.9 $\pm$ 4.8(stat) $\pm$ 2.9(syst) $\mevcc$. The product branching fraction is measured to be $\Br(\psp \ra \gamma \etacp) \times \Br(\etacp \ra \kskppp)$ = (7.03 $\pm$ 2.10(stat) $\pm$ 0.70(syst)) $\times$ $10^{-6}$.
This measurement complements a previous BESIII measurement of $\psp \ra \gamma\etacp$ with $\etacp \ra \kskp$ and $\kkp$.
\end{abstract}


\maketitle

\section{Introduction}

Compared to other charmonium states with masses below the open charm threshold, the properties of the $\etacp$ are not well-established.  The determination of the $\etacp$ mass, in particular, provides useful information about the spin-spin part of the charmonium potential.  The $\etacp$ was first observed at \emph{B}-factories~\cite{bell1,cleo1,babr1,babr2} and, to date, the only two measured branching fractions are for decays to $K\bar{K}\pi$ and $K^{+}K^{-}\pi^{+}\pi^{-}\pi^{0}$~\cite{pdg}. While the absolute branching fractions currently have poor precision, BaBar used the two-photon fusion process to measure the ratio of $\Br(\etacp \ra K^{+}K^{-}\pi^{+}\pi^{-}\pi^{0})$ to $\Br(\etacp \ra K_S^{0}K^{\pm}\pi^{\mp})$ to be 2.2 $\pm$ 0.5(stat) $\pm$ 0.5(syst)~\cite{babr3}.  The production of the $\etacp$ is also expected from magnetic dipole (M1) transitions~\cite{m1} of the $\psp$, and  $\psp \ra \gamma \etacp$ with $\etacp \ra K\bar{K}\pi$ has previously been observed by BESIII~\cite{kseff}. This analysis complements the previous analysis by focusing on the same radiative decay, $\psp \ra \gamma \etacp$, but with $\etacp \ra \kskppp$.

In our study, $\psp$ mesons are produced by the annihilation of electron-positron pairs at a center-of-mass energy of $3686$ MeV.  The production of the $\etacp$ through a radiative transition from the $\psp$ requires a charmed-quark spin-flip and, thus, proceeds via a M1 transition. Some of the generated $\etacp$ mesons will decay into hadrons, and then ultimately into detectable particles, like pions, kaons, and photons.  We study the decay exclusively by reconstructing the $\etacp$ from its hadronic decay products and analyze the $\etacp$ candidate mass for an evidence of $\psp \ra \gamma \etacp$. The experimental challenge of the measurement of this decay channel is to detect the 48~MeV radiative photons in an experimental environment with considerable backgrounds, therefore the success of this study depends on a careful and detailed analysis of all possible background sources.

\section{The experiment and data sets}

The data sample for this analysis consists of $1.06\times 10^{8}$
events produced at the peak of the $\psp$ resonance~\cite{npsp}.
Data were collected with an additional integrated luminosity of 42~pb$^{-1}$ at a
center-of-mass energy of $\sqrt{s}$=3.65~GeV to determine
non-resonant continuum background contributions. The data were
accumulated with the BESIII detector operated at the BEPCII $\EE$
collider.

The BESIII detector, described in detail in Ref.~\cite{bes3}, has an
effective geometrical acceptance of 93\% of 4$\pi$. It contains a
small cell helium-based main drift chamber (MDC) which provides
momentum measurements of charged particles; a time-of-flight system
(TOF) based on plastic scintillator which helps to identify charged
particles; an electromagnetic calorimeter (EMC) made of CsI (Tl)
crystals which is used to measure the energies of photons and provide
trigger signals; and a muon system (MUC) made of Resistive Plate
Chambers (RPC). The momentum resolution of the charged particles is $0.5$\% at
$1~\gevc$ in a 1~Tesla magnetic field. The energy loss
($dE/dx$) measurement provided by the MDC has a resolution
better than 6\% for electrons from Bhabha scattering. The photon
energy resolution can reach $2.5$\% ($5$\%) at $1~\gev$ in the barrel
(endcaps) of the EMC. And the time resolution of the TOF is $80$~ps in the
barrel and $110$~ps in the endcaps.

Monte Carlo~(MC) simulated events are used to determine the
detection efficiency, optimize the selection criteria, and study
the possible backgrounds. The simulation of the BESIII detector is based on
{\sc geant4}~\cite{geant4}, in which the interactions of the
particles with the detector material are simulated. The $\psp$
resonance is produced with \textsc{kkmc}~\cite{KKMC},
which is the event generator based on precise predictions of the Electroweak Standard Model
for the process $e^{+}  e^{-} \ra f\overline{f} + n\gamma$, where $f = e, \mu, \tau, d, u, s, c, b$, and $n$ is an integer number.
The subsequent decays are generated with {\sc EvtGen}~\cite{EvtGen}.
The study of the background is based on a sample of $10^8$ $\psp$
inclusive decays, generated with known branching
fractions taken from the Particle Data Group (PDG)~\cite{pdg}, or
with {\sc lundcharm}~\cite{Lundcharm} for the unmeasured decays.

\section{Event selection}
The decays of $\psp \ra \gamma \etacp$ with $\etacp \ra \kskppp$ are selected for this analysis.
A charged track should have good quality in the track fitting and be
within the angle coverage of the MDC, $|\cos\theta|<0.93$. A good
charged track (excluding those from $\ks$ decays) is required to pass
within 1~cm of the $\EE$ annihilation interaction point (IP)
in the transverse direction to the beam line and within 10~cm of the IP along the beam
axis. Charged-particle identification (PID) is based on combining the
$dE/dx$ and TOF information to the variable $\chi^{2}_{{\rm
PID}}(i)= (\frac{dE/dx_{\rm measured}-dE/dx_{\rm
expected}} {\sigma_{dE/dx}})^2+ (\frac{{\rm TOF}_{\rm
measured}-{\rm TOF}_{\rm expected}}{\sigma_{\rm TOF}})^2$. The values
$\chi^{2}_{{\rm PID}}(i)$ and the corresponding confidence levels
${\rm Prob_{PID}}(i)$ are calculated for each charged track for each
particle hypothesis $i$ (pion, kaon, or proton).

Photon candidates are required to have energy
greater than 25~$\mev$ in the EMC  both for the barrel region
($|\cos\theta|<0.8$) and the endcap region ($0.86<|\cos\theta|<0.92$).
In order to improve the reconstruction efficiency and the energy
resolution, the energy deposited in the nearby TOF counter is
included.
EMC timing requirements are used to suppress noise and remove energy deposits unrelated to the event. Candidate events must have exactly six charged tracks with net charge zero and at least one good photon.

$\ks$ candidates are reconstructed from secondary vertex fits to all the oppositely
charged-track pairs in an event (assuming the tracks to be $\pi^{\pm}$). The
combination with the best fit quality is kept for further analysis, where the $\ks$ candidate
must have an invariant mass within $10~\mevcc$ of the $\ks$ nominal
mass and the secondary vertex is well separated from the interaction point. At least one good $\ks$ is reconstructed, and the related information is used as input for the subsequent kinematic fit.

After tagging the $\pi^{+}\pi^{-}$ pair from the $\ks$, the other charged particles should be three pions and one kaon.
To decide the species of those particles, we make four different particle combination assumptions:
 $K^{+}\pi^{-}\pi^{+}\pi^{-}$, $\pi^{+}K^{-}\pi^{+}\pi^{-}$, $\pi^{+}\pi^{-}K^{+}\pi^{-}$, and $\pi^{+}\pi^{-}\pi^{+}K^{-}$.
For the different assumptions, four-momentum conservation constraints (4C) are required to be satisfied for each event candidate.
For each event, the M1-photon is selected with the minimum chi-square of the 4C kinematic fit ($\chi^{2}_{\rm 4C}$) by looping over all the good photons.
Then the $\chi^{2}_{\rm 4C}$ and the chi-squares of the particle-identification for kaon ($\chi^{2}_{K}$) and pions ($\chi^{2}_{\pi}$) are added together as the total chi-square ($\chi^{2}_{\rm total}$) for event selection. The types of particles are determined by choosing the smallest total chi-square.
Events with $\chi^{2}_{\rm total} < 60$ are accepted as the $\gamma \kskppp$ candidates.

To suppress the $\psp \to \pp\jpsi$, $\jpsi \ra \gamma \kskp$ decay, events are rejected if the recoil mass of any $\pp$ pair is within 15 $\mevcc$ of the $\jpsi$ nominal mass.
The  $\psp \to \eta\jpsi$, $\eta \to \gamma\pp$ events are rejected if the mass of $\kskp$ is greater than 3.05 $\gevcc$.
In order to suppress $\psp \to \etap \kskp$, $\etap \to \gamma \pp$ decays, events are removed if the mass of any $\gamma\pp$ combination is within 20 $\mevcc$ of the nominal $\etap$ mass.

\section{Data analysis}
The results of an analysis of the inclusive MC data sample showed that the primary source of background is $\psp \to \kskppp$. There are two mechanisms for this decay to produce background: a fake photon, or a photon from final-state radiation (FSR) is incorporated into the final state. Other backgrounds include $\psp \ra \pi^{0} \kskppp$ with a missing photon and initial state radiation (ISR). The phase space process $\psp \to \gamma \kskppp$ has the same final states as our signal, so it should be considered as an irreducible background.  As discussed in a later section, the size of this irreducible background is estimated using a region of $\kskppp$ mass away from the $\etacp$ mass.

\begin{figure*}[htbp]
\begin{center}
\includegraphics[width=3.0in,height=2.0in]
{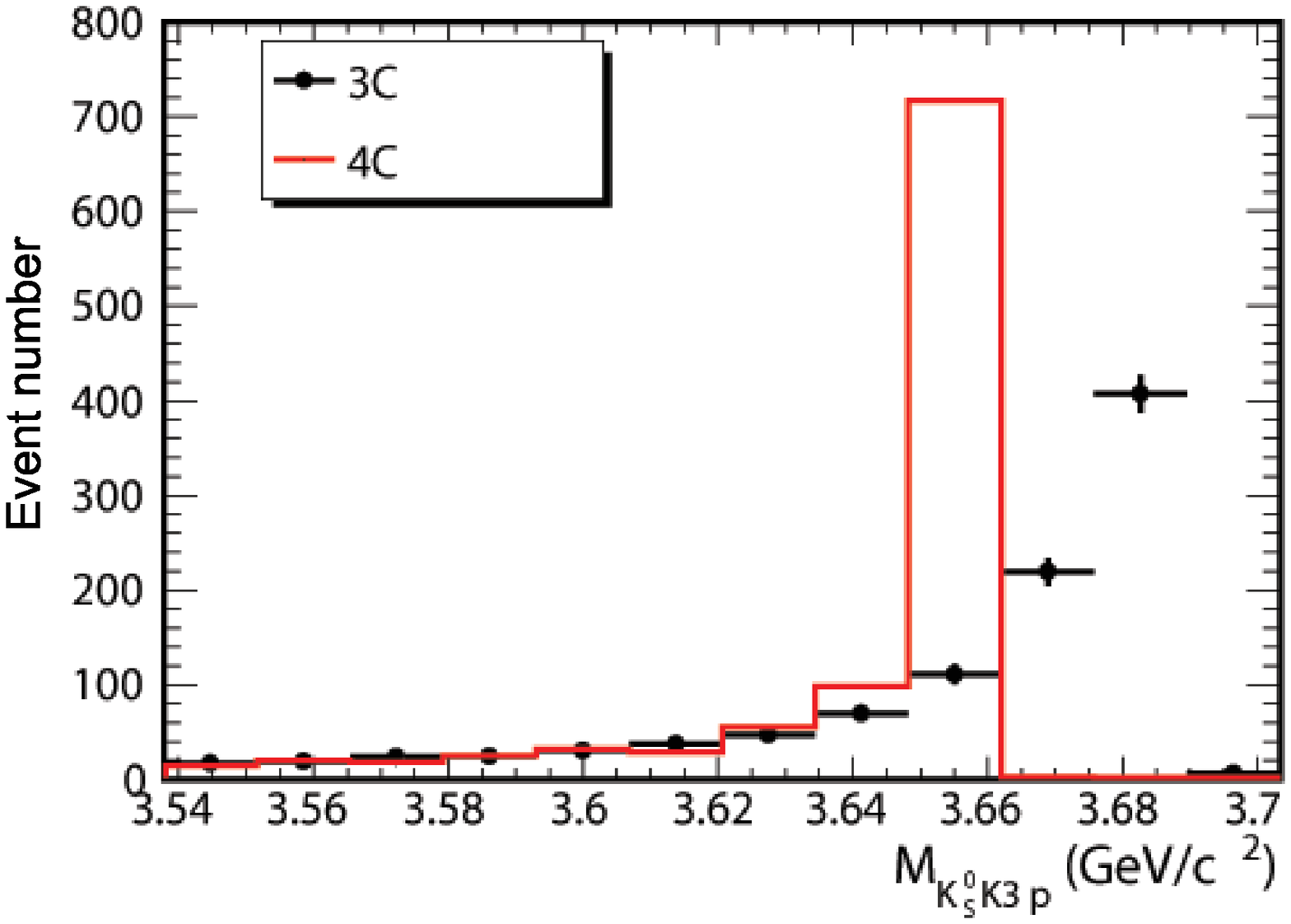}
\includegraphics[width=3.0in,height=2.0in]
{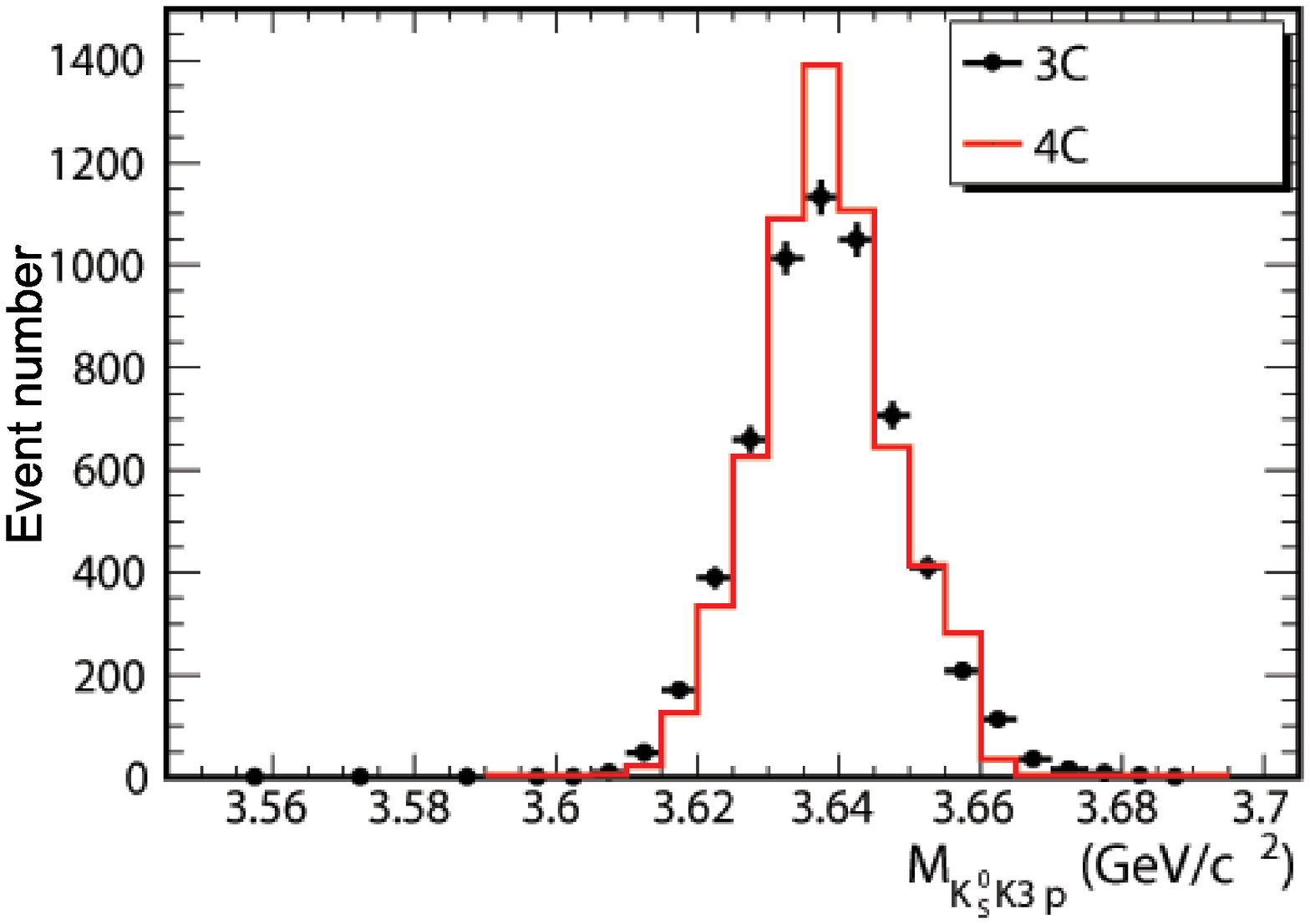}
\end{center}
\caption{Invariant mass spectrum of $\kskppp$ for the background $\psp \to \kskppp$ with a fake photon (left panel) and
the signal $\psp \to \gamma \etacp$, $\etacp \to \kskppp$ (right panel). The
points with error bars are 3C kinematic fit results, and the solid lines are 4C kinematic fit results.}
\label{fig:3c}
\end{figure*}

In the $\psp \to \kskppp$ background with a fake photon, a peak could be produced in the $\kskppp$ mass spectrum close to the expected $\etacp$ mass with a sharp cutoff due to the 25 MeV photon energy threshold. Considering that the fake photon does not contribute useful information to the kinematic fit, we set the photon energy free in the kinematic fit to avoid the mass distortion caused by the 25 MeV photon energy threshold. We call this the 3C kinematic fit and produce the mass spectrum based on it. MC studies demonstrate that with the 3C kinematic fit, the energy of the fake photon tends to zero, which is helpful in separating the signal from the fake photon background, as shown in Fig~\ref{fig:3c}~\cite{fsr}.

In the other $\psp \ra \kskppp$ background, a photon from final state radiation~($\gamma_{\rm FSR}$) could contaminate our signal. The $M^{\rm 3C}_{K_{S}^{0}K3\pi}$ with the FSR process has a long tail from $3.58~\gevcc$ to $3.68~\gevcc$ in our $\etacp$ signal region. We have to estimate the contribution of this FSR process, because it contributes to the background in our signal region and cannot be reduced for the same final states as the signal.
FSR is simulated in our MC generated data with {\sc PHOTOS}~\cite{photos}, and the FSR contribution is scaled by the ratio of FSR fractions in data and MC generated data for a control sample of $\psp \ra \gamma \pi^{+} \pi^{-} K^{+} K^{-}$ and $\psp \ra \gamma \pi^{+} \pi^{-} \pi^{+} \pi^{-}$~\cite{fsr}.
The background contributions from $\psp \ra \kskppp$ with fake photons and $\gamma_{\rm FSR}$ are estimated with MC distributions normalized according to branching ratios we measured.

\begin{figure*}[htbp]
\begin{center}
\includegraphics[width=5.0in,height=3.5in]
{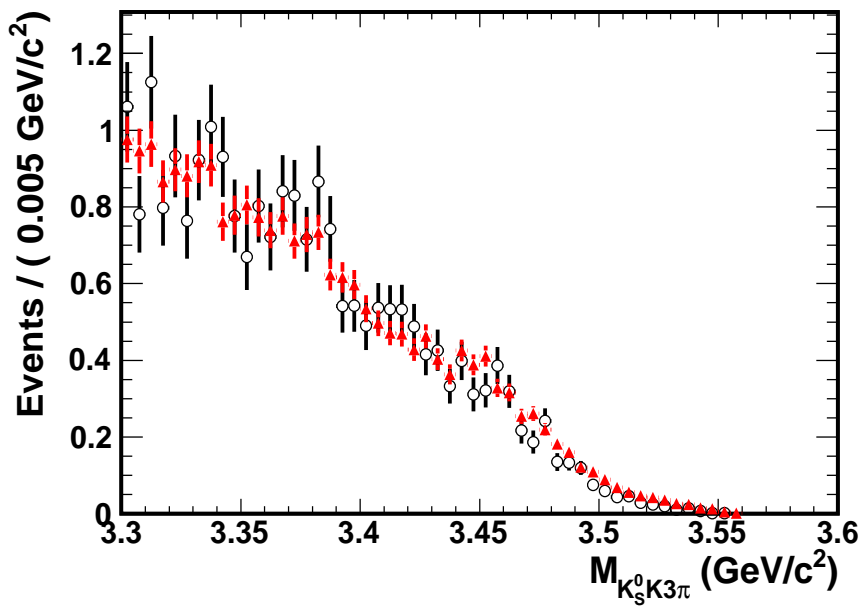}
\end{center}
\caption{The invariant mass distribution of $K_{S}^{0}K3\pi$ for the background from $\psp \ra \pi^{0} \kskppp$. The black circles with error bars show the background shape obtained from the collider data. The red triangles with error bars represent the $M_{K_{S}^{0}K3\pi}$ distribution from a corresponding MC sample.}
\label{fig:bk2_p3}
\end{figure*}

The channel $\psp \ra \pi^{0} \kskppp$ can contaminate our signal when one of the photons from the $\pi^{0}$ is not detected. MC generated events of the $\psp \ra \pi^{0} \kskppp$ process, based on the phase space model, and which satisfy the selection criteria for the $\psp \ra \gamma \kskppp$ signal, are taken to study this background and estimate its response. To prove the correctness of the MC simulation, the $\psp \ra \pi^{0} \kskppp$ control sample, which is selected from the colliding data, times the efficiency to reconstruct $\psp \ra \pi^{0} \kskppp$ events as $\psp \ra \gamma \kskppp$ is shown in Fig.~\ref{fig:bk2_p3} and compared with the same distribution obtained from the corresponding $\psp \ra \pi^{0} \kskppp$ MC simulation. The consistency of the two distributions is checked by the Kolmogorov-Smirnov test~\cite{kol}, and a good agreement is verified (the consistency probability reaches 0.28).

The background from the continuum (including ISR) is estimated with collider data taken at a center of mass energy of $3.65~\gev$.  The events must pass the signal selection requirements and are then normalized according to differences in integrated luminosity and cross section.  Particle momenta and energies are scaled to account for the beam-energy difference. The resultant number and the $K_{S}^{0}K3\pi$ invariant mass shape considering these scale factors ($f_{\rm continuum}$ = 3.6) are used in the final fit.

The background from phase space has the same final states as the signal. To select a clean phase space sample, the $M_{K_{S}^{0}K3\pi}$ region $[3.20, 3.30]~\gevcc$ is chosen. This choice is made because there is a long $\eta_{c}$ tail in the area $M_{K_{S}^{0}K3\pi} < 3.0~\gevcc$ which originates from the decay channel $\psp \ra \gamma \eta_{c}$.  There are three obvious peaks in the area $M_{K_{S}^{0}K3\pi} >3.3~\gevcc$ which are from the decay channel $\psp \ra \gamma \chi_{cJ}, (J = 0, 1,{\rm and}~2)$.  The branching fraction of the phase space process is calculated to be $1.73\times 10^{-4}$. The $K_{S}^{0}K3\pi$ invariant mass spectrum of MC phase space events is used in the final fit, while the number of events is left floating. The number of phase space events obtained by fitting the mass spectrum is consistent with that estimated by the branching fraction we calculated.

\begin{figure*}[htbp]
\begin{center}
\includegraphics[width=5.0in,height=3.5in]
{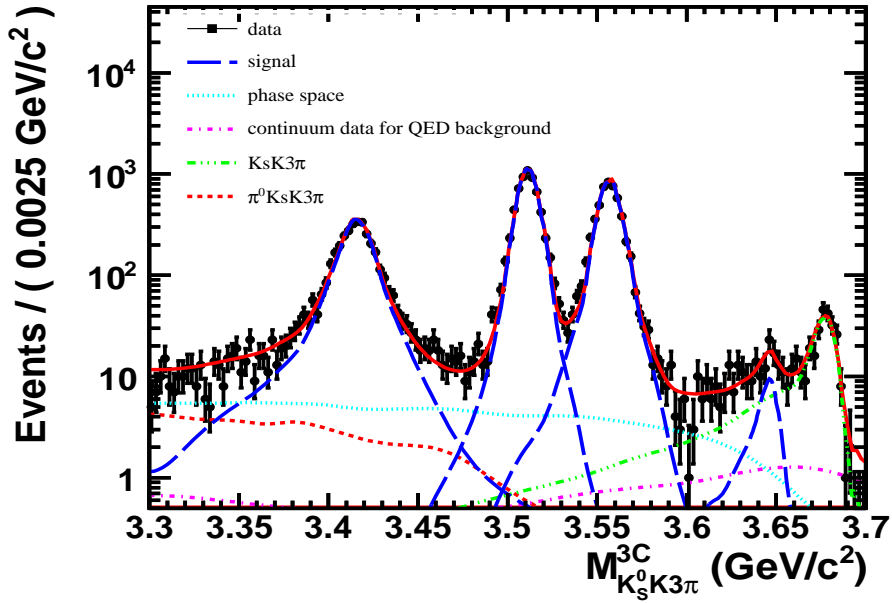}
\end{center}
\caption{The results of fitting the mass spectrum for $\chi_{cJ}$ and $\etacp$. The black dots are the collider data, the blue long-dashed line shows the $\chi_{cJ}$ and $\etacp$ signal shapes, the cyan dotted line represents the phase space contribution, the violet dash-dotted line shows the continuum data contribution, the green dash-double-dotted line shows the contribution of $\psp \ra \kskppp$, and the red dashed line is the contribution of $\psp \ra \pi^{0} \kskppp$.}
\label{fig:fitting_total}
\end{figure*}

In the $K_{S}^{0}K3\pi$ mass spectrum fitting, the fitting range is from $3.30~\gevcc$ to $3.70~\gevcc$ so that the contributions of backgrounds and $\chi_{cJ}(J= 0, 1, {\rm and} ~2)$ can be taken into account. The final mass spectrum and the fitting results are shown in Fig.~\ref{fig:fitting_total}. The fitting function consists of the following components: $\etacp$, $\chi_{cJ}(J= 0, 1, {\rm and}~2)$ signals and $\psp \ra \kskppp$, $\psp \ra \pi^{0} \kskppp$, ISR, and phase space backgrounds.
The line shapes for $\chi_{cJ}$ are obtained from MC simulations. These can describe the $\chi_{cJ}$ spectrum well in the collider data after applying the kinematic fit correction~\cite{guoyp}.
The line shape for $\etacp$ produced by such a M1 transition is given by:
\begin{equation}
(E^{3}_{\gamma} \times BW(m) \times damping(E_{\gamma})) \otimes Gauss(0,\sigma),
\label{fun:etacp}
\end{equation}
where $BW(m)$ is the Breit-Wigner function,  $m$ is the invariant mass of $K_S^{0}K3\pi$, $E_{\gamma} = \frac{m^{2}_{\psp} - m^{2}}{2m_{\psp}}$ is the energy of the transition photon in the rest frame of $\psp$, $damping(E_{\gamma})$ is the function to damp the diverging tail raised by $E^{3}_{\gamma}$ and $Gauss(0, \sigma)$ is the Gaussian function describing the detector resolution. The detector resolution is determined by the MC study, and the difference of data and MC has been taken into account which introduces negligible uncertainties in branching fraction, mass and width measurements comparing with other factors. 
The form of the damping function is somewhat arbitrary, and one suitable function used by KEDR~\cite{KDER} for a similar process is
\begin{equation}
damping(E_{\gamma}) = \frac{E^{2}_{0}}{E_{\gamma}E_{0} + (E_{\gamma} - E_{0})^{2}},
\label{fun:kder}
\end{equation}
where $E_{0} = \frac{m^{2}_{\psp} - m^{2}_{\etacp}}{2m_{\psp}}$ is the peaking energy of the transition photon. Another damping function used by CLEO~\cite{cleo} is inspired by the overlap of wave functions
\begin{equation}
damping(E_{\gamma}) = \exp(-E^{2}_{\gamma}/8\beta^{2}),
\label{fun:cleo}
\end{equation}
with $\beta = (65.0 \pm 2.5$) MeV from CLEO's fit. In our analysis, the KEDR function (Eq.~\ref{fun:kder}) is used in the fitting to give the final results, and the CLEO one (Eq.~\ref{fun:cleo}) is used to estimate the possible uncertainty caused by the form of damping functions.

The result for the yield of $\etacp$ events is $57\pm17$ with a significance of 4.2$\sigma$. The significance is calculated from log-likelihood differences between fits with and without the $\etacp$ component. The robustness of this result was tested by considering different damping factor forms, FSR fractions, and background assumptions. In all the cases, the statistical significance is found to be larger than 4$\sigma$.
The resulting mass and width from the fit are $3646.9\pm1.6$ $\mevcc$ and $9.9\pm4.8$ $\mevcc$ (statistical errors only), respectively. We find the product branching fraction $\Br(\psp \ra \gamma \etacp)\times \Br(\etacp \ra \kskppp) = (7.03~\pm~2.10)~\times~10^{-6}$ with the efficiency of $11.1\%$ for the signal selection.

\section{Estimation of systematic uncertainties}

The systematic uncertainties in the $\etacp$ mass and width measurements are estimated by the uncertainties in the damping factor, scale factor and the number of $\psp\to\pi^{0}\kskppp$ events. The results are summarized in Table~\ref{table:sys_mass_width}, and described in more detail in the following.

\begin{table*}[htbp]
\caption{Uncertainties in the mass and width of $\etacp$.}\label{table:sys_mass_width}
\begin{center}
\begin{tabular}{c  c  c }
  \hline
  Source & mass uncertainty & width uncertainty \\
  \hline
  \hline
 Damping factor & $<0.1\%$ & $28\%$\\
 Scale factor & negligible & $5\%$\\
 No. of $\piz \kskppp$ & $<0.1\%$ & $5\%$\\
 \hline
 Total & $<0.1\%$ & $29\%$\\
\hline
\end{tabular}
\end{center}
\end{table*}

We change the damping factor to the CLEO form, then compare the results with that obtained with the KEDR form, and the difference is taken as the uncertainty originating from the  damping factor.
The background shape of $\psp \ra \kskppp$ could influence the fitting results, so we change the FSR scale factor of 1.46 by 1$\sigma$ to $1.412$ and $1.504$, and
the difference in the results is taken as the uncertainty coming from scale factor.
In the fitting of the mass spectrum, the number of events for $\psp \to \piz \kskppp$ is fixed. We change the number of events by $1\sigma$, and take the difference in the results as the uncertainty originating from the number of background events from $\psp \to \piz \kskppp$ events.

The systematic errors in the measurement of the branching fraction are summarized in Table~\ref{table::sys error} and explained below.

\begin{table*}[htbp]
\begin{center}
\parbox{0.8\textwidth}{\caption{Summary of systematic uncertainties in the measurement of $\Br(\psp \ra \gamma \etacp,  \etacp \ra  \kskppp)$ .}\label{table::sys error}}
\newline
\begin{center}
\begin{tabular}{c c}
\hline
Sources                          &         Systematic uncertainties  \\
\hline
\hline
MDC tracking                     &         4\%    \\
Photon reconstruction            &         1\%    \\
$K_S^{0}$ reconstruction         &         4\%    \\
Kinematic fitting and PID        &         2\%    \\
Total number of $\psp$           &         0.8\%  \\
Damping factor                   &         2\%    \\
Scale factor                     &         5\%    \\
No. of $\psp\to\piz\kskppp$      &         2\%    \\
$\etacp$ width                   &         3\%    \\
Intermediate states              &         5\%    \\
\hline
Total                            &         10\%   \\\hline
\end{tabular}
\end{center}
\end{center}
\end{table*}

The tracking efficiencies for $K^{\pm}$ and $\pi^{\pm}$ as functions of transverse momentum have been studied with the process $\jpsi\to\kskp,K_{S}^{0}\to\pp$ and $\psp\to\pp\jpsi$, respectively. The efficiency difference between data and MC is $1\%$ for each $K^{\pm}$ track or $\pi^{\pm}$ track \cite{kaoneffi, pioneffi}. So the uncertainty of the tracking efficiency is $4\%$ for four charged tracks. The uncertainty of the two pions from $\ks$ is not included here, because it is included in the $\ks$ uncertainty.

The uncertainty due to photon reconstruction is $1\%$ per photon~\cite{photoneffi}. This is determined from studies of photon detection efficiencies in the process $\jpsi\to\rho^{0}\piz$, $\rho^{0}\to\pp$ and $\piz\to\gamma\gamma$.

Three parts contribute to the efficiency for $K_{S}^{0}$ reconstruction: the geometric acceptance, tracking efficiency and the efficiency of $K_{S}^{0}$ selection. The first part was estimated using an MC sample, and the other two were studied by the process $\jpsi\to K^{*}\bar{K^{0}}+c.c.$. The difference between data and MC is estimated to be $4\%$.

\begin{figure*}[htbp]
\begin{center}
\includegraphics[width=3.0in,height=2.0in]
{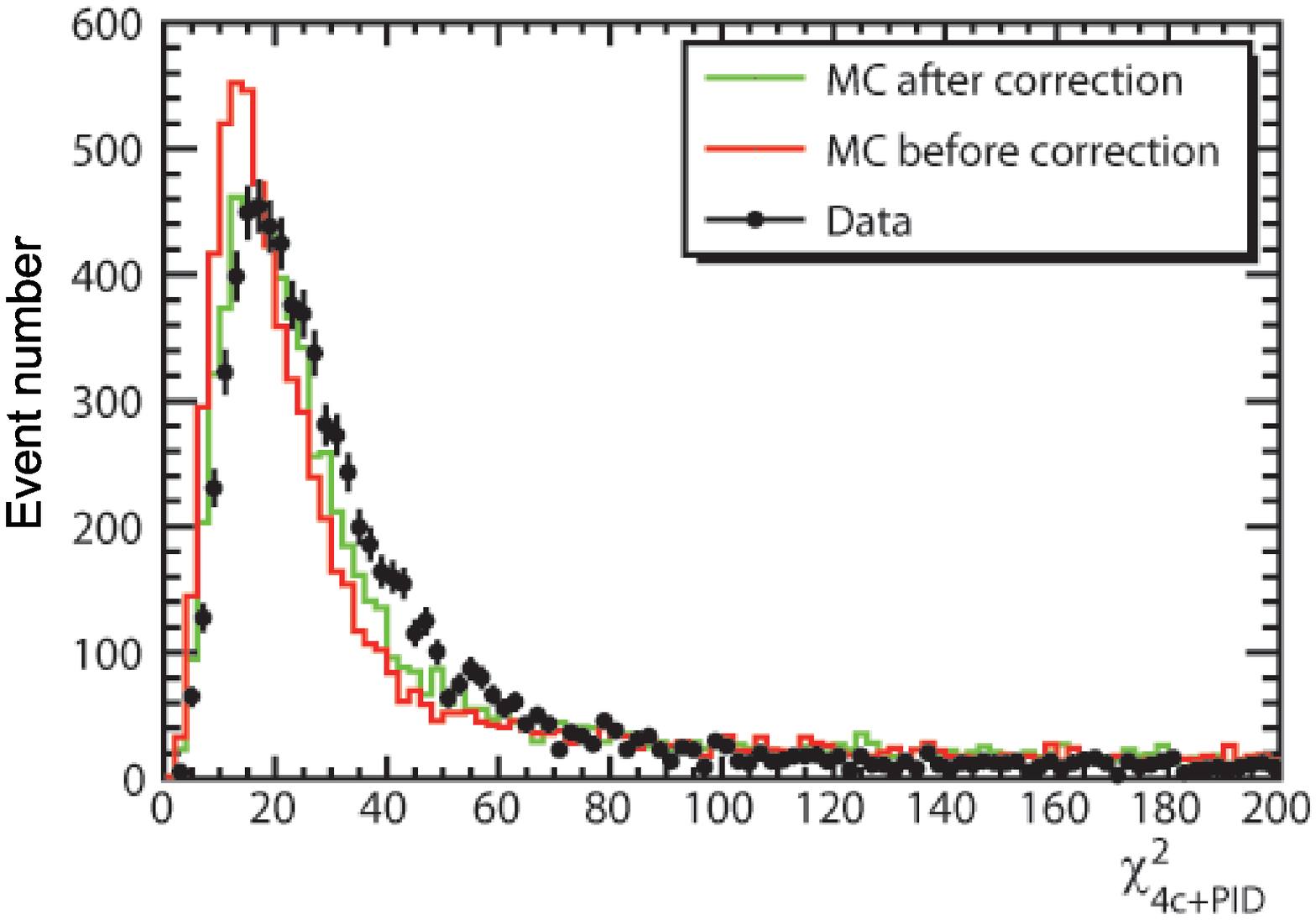}
\includegraphics[width=3.0in,height=2.0in]
{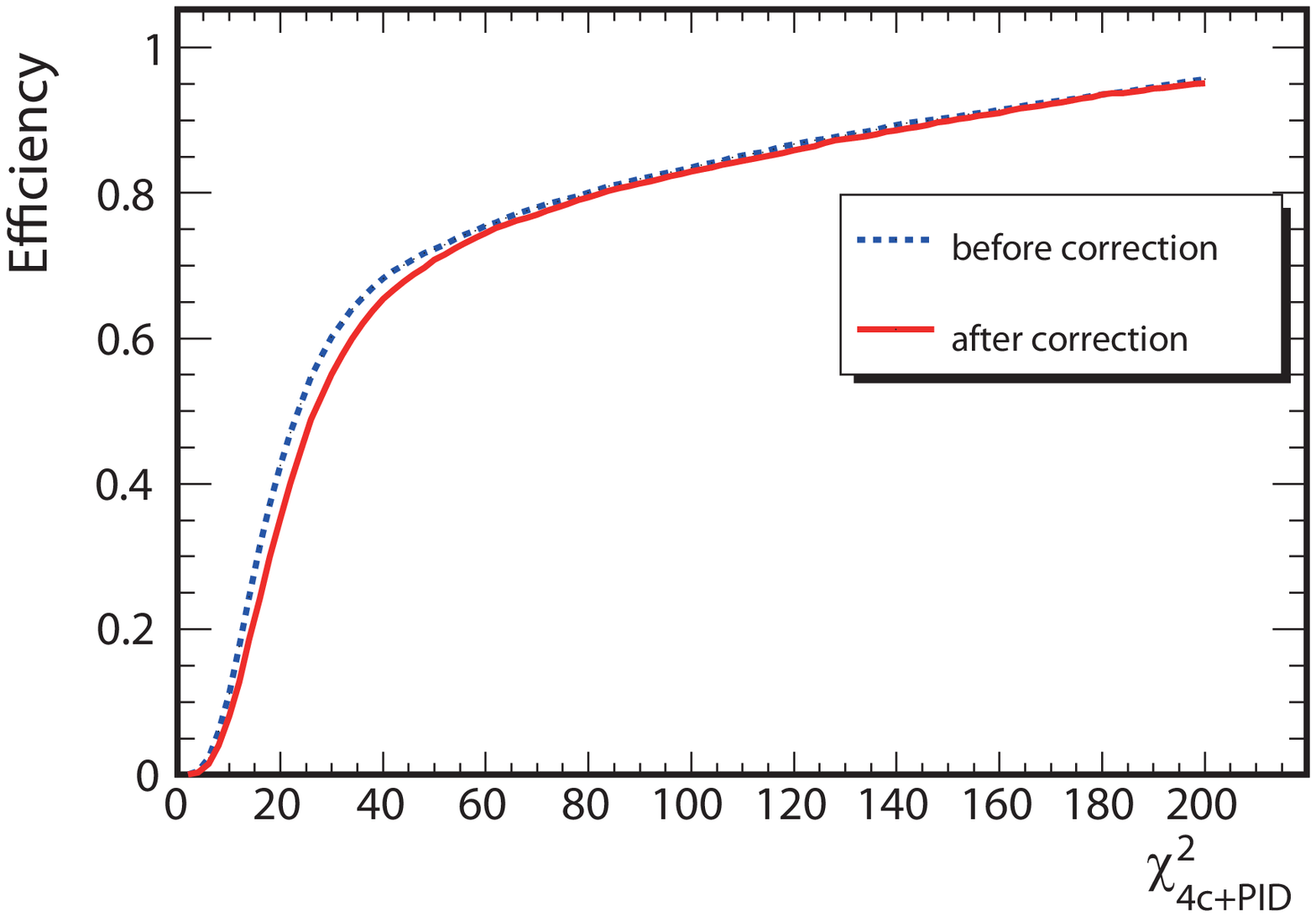}
\end{center}
\caption{[left panel]The $\chi^{2}_{\rm 4C+PID}$ distribution with and without the correction in MC and in data. The black dots show the distribution of $\chi^{2}_{\rm 4C+PID}$ in the data, the orange (green) histogram represents the distributions of $\chi^{2}_{\rm 4C+PID}$ without (with) correction in MC.
 [right panel] Efficiency results with and without correction at different $\chi^{2}_{\rm 4C+PID}$ cuts.}
\label{fig:sys_kinematicfit}
\end{figure*}

To estimate the uncertainty of kinematic fitting, we first correct the track helix parameters ($\phi_{0}$, $\kappa$, $tg\lambda$) to reduce the difference on $\chi^{2}_{\rm 4C}$ from kinematic fitting between data and MC, where $\phi_{0}$ is the azimuthal angle specifies the pivot with respect to the helix center, $\kappa$ is the reciprocal of the transverse momentum and $tg\lambda$ is the slope of the track. The correction factors are obtained from $\jpsi\to\phi f_{0}(980)$, $\phi\to\kk$ and $f_{0}(980)\to\pp$. The MC samples after correction are used to estimate the efficiency and fit the invariant mass spectrum.
Fig.~\ref{fig:sys_kinematicfit} (left) shows the $\chi^{2}_{\rm 4C+PID}$ distribution with and without the correction in MC and in data.  The distribution of $\chi^{2}_{\rm 4C+PID}$ with correction is closer to the data than without correction. However, the agreement is not perfect, and we take the systematic uncertainty to be the difference of the efficiency between MC before and after correction~\cite{guoyp}. The comparison is shown in Fig.~\ref{fig:sys_kinematicfit} (right). The systematic uncertainty from kinematic fitting is $2\%$ with $\chi^{2}_{\rm 4C+PID} < 60$.

We also change the form of the damping factor, the value of the FSR scale factor and the number of events for $\psp \to \piz \kskppp$ to estimate the uncertainties in the branching fraction, which is the same as the
method to estimate the uncertainties of $\etacp$ mass and width.
The total number of $\psp$ events is estimated by the inclusive hadronic events, and the uncertainty is $0.8\%$~\cite{npsp}.

To estimate the uncertainty due to the $\etacp$ width,  we change the $\etacp$ width of 9.9 $\mevcc$ by 1$\sigma$ to 5.1 $\mevcc$ and 14.7 $\mevcc$ in the
   MC simulation. Comparing the efficiencies with 11.1\%, which is used in calculating the branching fraction, we find a difference of 3\%.

 For the uncertainty from intermediate states, we generate MC samples including these states ($K^{*}(892)$, $\rho$) and compare the corresponding efficiencies.
We take the $5\%$ difference as the uncertainty.

We assume that all the sources of systematic uncertainties are independent and the overall systematic uncertainties are obtained by adding all single ones in quadrature.

\section{Conclusion}
We observe the decay mode $\etacp \ra \kskppp$ and establish the M1 transition of $\psp \ra \gamma \etacp$ using this decay mode.
The mass of the $\etacp$ is measured to be 3646.9 $\pm$ 1.6(stat) $\pm$ 3.6(syst) $\mevcc$, and the width is 9.2 $\pm$ 4.8(stat) $\pm$ 2.9(syst) $\mev$. 
Comparing with BESIII previous measurements~\cite{kseff}, the width is consistent with each other within 1 standard deviation and the mass is about 2 standard deviation.  The product branching fraction is measured to be $\Br(\psp \ra \gamma \etacp) \times \Br(\etacp \ra \kskppp)$ = (7.03 $\pm$ 2.10(stat) $\pm$ 0.70(syst)) $\times$ $10^{-6}$. The  statistical significance is greater than 4 standard deviation.

To compare with the BABAR results~\cite{babr3},
\begin{equation}
\frac{\Br(\etacp \ra K^{+}K^{-}\pi^{+}\pi^{-}\pi^{0})}{\Br(\etacp \ra K_S^{0}K^{\pm}\pi^{\mp})} = 2.2\pm0.5\pm0.5,
\end{equation}
we take the value of (4.31 $\pm$ 0.75) $\times$ $10^{-6}$ as measured by BESIII for $\Br(\psp \ra \gamma \etacp) \times \Br(\etacp \ra K_S^{0}K^{\pm}\pi^{\mp})$~\cite{kseff}, and assuming that
\begin{equation}
\frac{\Br(\etacp \ra K^{+}K^{-}\pi^{+}\pi^{-}\pi^{0})}{\Br(\etacp \ra \kskppp)}  = 1.52,
\end{equation}
 where the value 1.52 is calculated in $\chi_{cJ}$ decays, which has the same isospin,
we obtain
\begin{equation}
\frac{\Br(\etacp \ra K^{+}K^{-}\pi^{+}\pi^{-}\pi^{0})}{\Br(\etacp \ra K_S^{0}K^{\pm}\pi^{\mp})} = 1.52 \cdot \frac{\Br(\etacp \ra \kskppp)}{\Br(\etacp \ra K_S^{0}K^{\pm}\pi^{\mp})}  = 2.48\pm0.56\pm0.33.
\end{equation}
These two results are consistent with each other after considering the statistical and systematic uncertainties.

\acknowledgments
The BESIII collaboration is grateful to the staff of BEPCII and the computing center for their tireless efforts.
This work is supported in part by the Ministry of Science and Technology of
China under Contract No. 2009CB825200; National Natural Science Foundation
of China (NSFC) under Contracts Nos. 10625524, 10821063, 10825524,
10835001, 10935007, 11125525, 11235011, 10979038, 11079030, 11005109, 11275189, U1232201; Joint Funds of the National Natural Science
Foundation of China under Contracts Nos. 11079008, 11179007; the Chinese Academy
of Sciences (CAS) Large-Scale Scientific Facility Program; CAS under Contracts
Nos. KJCX2-YW-N29, KJCX2-YW-N45; 100 Talents Program of CAS; the Fundamental Research Funds for
the Central Universities under Contracts No. 2030040126, China; German Research Foundation DFG under Contract No. Collaborative Research Center CRC-1044;
Istituto Nazionale di Fisica Nucleare, Italy; Ministry of Development of Turkey under Contract No.
DPT2006K-120470; U. S. Department of Energy under Contracts Nos. DE-FG02-04ER41291, DE-FG02-94ER40823, DE-FG02-05ER41374; U.S. National Science Foundation; University
of Groningen (RuG); the Helmholtzzentrum f{\"u}r Schwerionenforschung GmbH (GSI),
Darmstadt; and WCU Program of National Research Foundation of Korea under Contract No. R32-2008-000-10155-0.

\end{document}